\documentclass[preprint]{aastex}

\slugcomment{Submitted to Astronomical Journal}

\shorttitle{VLA Serpens sources}
\shortauthors{Eiroa et al.}

\begin{document}


\title{ VLA 3.5 cm continuum sources in the Serpens cloud core
    }


\author{C. Eiroa}
\affil{Dpto. F\'\i sica Te\'orica, C-XI, Facultad de Ciencias, 
Universidad Aut\'onoma de Madrid, Cantoblanco, 28049 Madrid, Spain }
\email{carlos.eiroa@uam.es}
\author{J.M Torrelles}
\affil{Instituto de Ciencias del Espacio (CSIC) and Institut d'Estudis 
Espacials de Catalunya, C/ Gran Capit\'a 2-4, 08034 Barcelona, Spain. On 
sabbatical leave at the UK Astronomy Technology Centre. Royal Observatory 
Edinburgh}
\email{torrelles@ieec.fcr.es}
\author{S. Curiel}
\affil{Instituto de Astronom\'\i a, UNAM, Apdo. Postal 70-264, 04510 M\'exico 
D.F., M\'exico and Harvard-Smithsonian Center for Astrophysics, 60 Garden St, 
Cambridge, MA02138, USA}
\email{scuriel@astroscu.unam.mx}
\and
\author{A.A. Djupvik}
\affil{Nordic Optical Telescope, Apdo. 474, 38700 Santa Cruz de La Palma}
\email{akaas@not.iac.es}



\begin{abstract}
We present VLA 3.5 cm continuum observations of the Serpens cloud core. 
22 radio 
continuum sources are detected. 16 out of the 22 cm sources are suggested to 
be associated with young stellar objects (Class 0, Class I, flat-spectrum, 
and Class II) of the young Serpens cluster. The rest of the VLA sources 
plausibly are background objects. Most of the Serpens cm sources likely 
represent thermal radio jets; on the other hand, the radio continuum emission
of some sources could be due to a gyrosynchroton mechanism arising from
coronally active young stars. The Serpens VLA sources are spatially 
distributed into two groups; one of them located towards the NW clump of the 
Serpens core, where only Class 0 and Class I protostars are found to present 
cm emission, and a second group located towards the SE clump, where radio 
continuum sources are associated with objects in evolutionary classes from 
Class 0 to Class II. This subgrouping is similar to that found in the 
near IR, mid-IR and mm wavelength regimes.   
\end{abstract}





\section{Introduction}

The Serpens cloud core (size $\sim 6' $ in diameter) is one of the more active,
nearby star-forming regions which has been the subject of many observational 
studies during the last $\sim$ 15 years. The core is populated by a dense and 
extremely young cluster whose members are found in many different evolutionary 
stages. Near-IR surveys have identified more than 150 near-IR sources embedded 
in the cloud core (Eiroa \& Casali, 1992; Sogawa 
et al. 1997; Giovannetti et al. 1998; Kaas 1999). Many of these sources have 
been classified as Class II young stellar objects (YSOs) - i.e. embedded T 
Tauri stars - by means of mid-IR ISO observations (Kaas et al. 2004); these 
observations also reveal a large number of flat-spectrum sources and Class I 
protostars. Submillimeter, millimeter, and far-IR observations show that the 
cloud core is also populated by many Class 0 protostars and protostellar 
condensations (Casali et al. 1993; Hurt \& Barsony 1996; Testi \& Sargent 
1998). Line profiles of different molecules and transitions suggest infall 
motions of the gas associated with some of the Class 0 protostars (e.g Hurt 
et al. 1996; Gregersen et al., 1997; Mardones et al. 1997; Wolf-Chase et al. 
1998; Hogerheidijde et al. 1999; Williams \& Myers 2000; Narayanan et al. 
2002). X-ray sources have  been detected and identified with some of the 
YSOs (Preibisch 1998, 2003, 2004). In addition, an embedded young brown dwarf 
has been identified in the Serpens cloud (Lodieu et al. 2002). Further 
signspots related to star formation activity are found in the Serpens
core: HH objects (e.g Davis et al. 1999; Ziener \& Eisl\"offel 1999), H$_2$ 
emission (e.g. Eiroa et al. 1997; Herbst et al. 1997; Hodapp 1999), and 
molecular outflows (White et al. 1995; Wolf-Chase et al; 1998; Williams \& 
Myers 2000). Thus, the Serpens core represents an excellent laboratory for 
studying physical processes associated with star formation and their mutual 
inter-relations.

Very little has been done, however, in the radio continuum. Such observations 
are relevant because, among other aspects, they precisely reveal the positions 
of YSO objects, trace stellar winds at positions very close to the driving
sources, and  provide insights into thermal and synchroton emission in YSOs, 
stellar coronal activity of YSOs and magnetic fields. Radio continuum 
observations of Serpens were reported by Rodr\'{\i}guez et al. (1980) and 
Snell \& Bally (1986). Rodr\'{\i}guez et al. (1989) and Curiel et al. (1993) 
studied a somehow unique triple radio-source with thermal and non-thermal 
emission components proposed to be a radio jet precursor of Herbig-Haro 
objects. This radio source is associated with the Class 0/Class I protostar 
smm 1/FIRS 1, the most  luminous object embedded in the cloud core (Harvey et 
al. 1984; Casali et al. 1993). Smith et al. (1999) has analysed some VLA 
data of the radio emission coming from the SVS 4 region (Strom et al. 1976; 
Eiroa \& Casali 1989). However, a detailed analysis of the radio emission in 
the Serpens cloud core is still lacking.

In this work we present VLA 3.5 cm continuum observations carried out by us 
toward the Serpens cloud. Our aim is to make a first approach to the Serpens 
cluster behaviour at radio continuum wavelengths, filling at least partly  
the existent gap at these frequencies. The layout of the paper is as follows. 
In Section ~\ref{Observations} we present details of the VLA observations.
Section ~\ref{Results} presents the VLA sources detected in the data; in that 
section we suggest associations of the radio sources with Serpens YSOs identified in other wavelengths regimes. Section ~\ref{Discussion} presents a 
discussion on the plausible nature of the radio sources and their clustering 
in the overall frame of the Serpens YSO cluster. Section \ref{Conclusions} 
presents our conclusions. Finally, a short description of the individual radio
sources and their counterparts together with their corresponding positions are
given in the appendix (Section ~\ref{Individual}). 

\section{Observations}
\label{Observations}
\begin{table*}[ht]
\caption{VLA observational parameters at 3.5 cm continuum}
\begin{tabular}{cllccc}
\hline		 		
VLA configuration & Phase Center & & Date & Beam size & rms \\
      &$\alpha$(J2000.0)&$\delta$(J2000.0) & &$''$;PA$^{\circ}$&
$\mu$Jy/beam\\
\hline
C/D &18h29m57.8s& 1$^{\circ}$ 14$'$ 07.2$''$ &24/10/1993 &7.5$\times$3.1; 81&15\\
D   &18h29m49.8s& 1$^{\circ}$ 15$'$ 20.8$''$ &06/02/1994 &9.8$\times$7.8; 37&15\\
\hline
\end{tabular}
\label{radioparameters}
\end{table*}


Observations at 3.5 cm continuum where made with the Very Large Array (VLA)
of the National Radio Astronomy Observatory (NRAO)\footnote{NRAO is a facility 
of the National Science Foundation operated under cooperative agreement by
Associated Universities Inc.} in its C/D and D arrays during 1993 October 24th 
and 1994 February 6th, respectively. The phase center of the interferometer was
at the position of the infrared source SVS 20 (Eiroa \& Casali, 1992) and at 
the position of the Serpens thermal radio jet (Curiel et al. 1993) for the C/D 
and D configurations, respectively (Table \ref{radioparameters}). The 
observations were made in both circular polarizations with an effective 
bandwidth of 100 MHz. The absolute amplitude calibrator was 1328+307, with an 
adopted flux density of 5.27 Jy, while 1749+096 was used as phase calibrator. 
The data were edited, calibrated and imaged following standard procedures 
using the NRAO AIPS package. In order to measure the flux density of the 
individual radio sources, We produced cleaned maps setting the ROBUST 
parameter of the AIPS task IMAGR to 0 to optimize the tradeoff between angular 
resolution and sensitivity. The resulting synthesized beam sizes and rms 
sensitivities of the maps are given in Table \ref{radioparameters}.

\section{Results}
\label{Results}
\clearpage

\begin{figure}[ht]
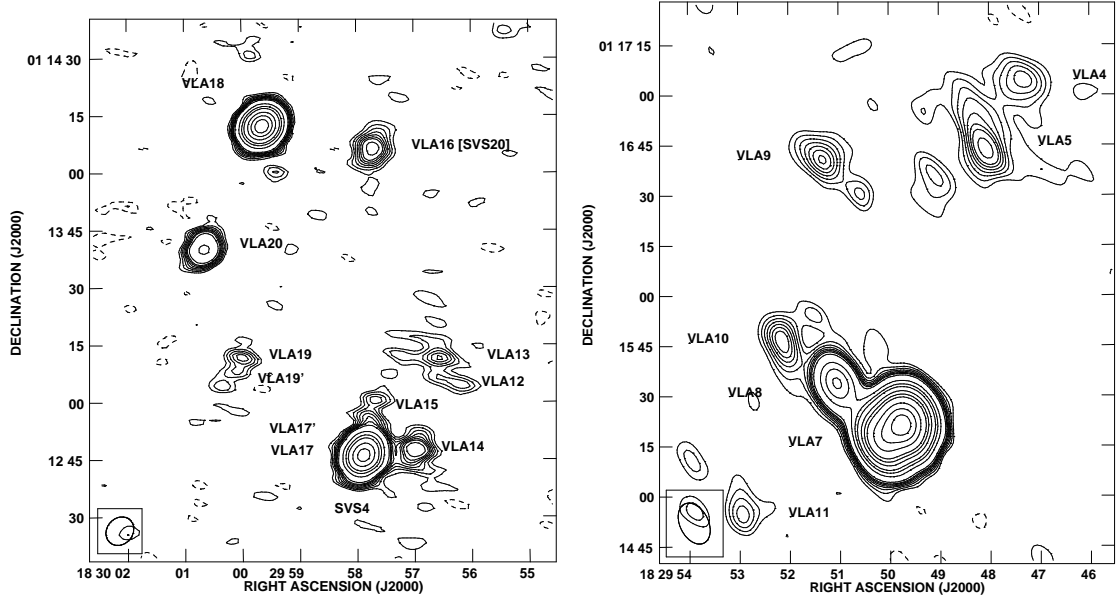

\scalebox{0.40}{\includegraphics*[40,120][570,670]{f1a.ps}}
\scalebox{0.39}{\includegraphics*[40,105][570,685]{f1b.ps}}
\caption{3.5 cm continuum contour map of the Serpens cloud core. The detected 
radio sources (Table 2) are labelled {\it (a)} Left: the SVS 20 field  made 
with natural weighting in order to show up some weak extended emission around 
some of the 3.5 cm continuum sources. Contour levels are -5, -4, -3, -2, 2, 
3, 4, 5, 6, 7, 8, 10, 20, 30,
40, 60, 80, 100 $\times$ 15~$\mu$Jy/beam, the rms noise of the map. The 
half-power contour of the beam (7.9$''$$\times$6.8$''$) is shown in the bottom 
left corner. The association of the binary near-IR source SVS 20 and the 
position of the SVS 4 near-IR source group  are indicated. {\it (b)} Right: Same as (a) but for the field around the Serpens 
smm 1/ FIRS 1/ triple radio continuum source (labelled as VLA~7). Contour 
levels are -5, -4, -3, -2, 2, 3, 4, 5, 6, 7, 8, 10, 20, 30, 40, 60, 80, 100, 
150, 200, 250 $\times$ 15~$\mu$Jy/beam, the rms noise of the map. The 
half-power contour of the beam (13.0$''$$\times$8.5$''$) is shown in the 
bottom left corner.}
\label{svs20_radio}
\end{figure}    
\clearpage

\begin{table*}
\caption{Equatorial coordinates (J2000.0) and flux densities of 
Serpens VLA 3.5 cm  continuum sources$^a$. The last column indicates 
whether the cm sources have  plausible counterparts at other wavelengths (Y), 
uncertain (?) or not (N)}
\begin{tabular}{lclllll}
\hline		 		
Source & Configuration$^b$ & $\alpha$ &$\delta$ &Peak & Flux density& 
Association\\
    & &         &            & (mJy/beam)     &   (mJy)  &       \\
\hline			      
1  & D & 18h29m 34.19s & 1$^{\circ}$ 15' 10.7''& 0.56 & 0.56 &?\\
2  & D & 18h29m 35.00s & 1$^{\circ}$ 15' 03.7''& 2.05 & 2.17 &?\\
3  & D & 18h29m 44.05s & 1$^{\circ}$ 19' 22.4''& 2.10 & 2.10 &N  \\
4  & D & 18h29m 47.29s & 1$^{\circ}$ 17' 04.6''& 0.14 & 0.14 &?\\
5  & D & 18h29m 48.10s & 1$^{\circ}$ 16' 44.7''& 0.21 & 0.28 &Y \\
6  & D & 18h29m 49.51s & 1$^{\circ}$ 19' 56.8''& 1.87 & 1.87 &N  \\
7  & D & 18h29m 49.79s & 1$^{\circ}$ 15' 20.8''& 4.29 & 7.54 &Y  \\
8  & D & 18h29m 51.06s & 1$^{\circ}$ 15' 33.9''& 0.64 & 0.64 &N  \\
9  & D & 18h29m 51.30s & 1$^{\circ}$ 16' 40.9''& 0.15 & 0.15 &Y  \\
10 & D & 18h29m 52.12s & 1$^{\circ}$ 15' 46.0''& 0.14 & 0.14 &Y  \\
11 & D & 18h29m 52.94s & 1$^{\circ}$ 14' 54.0''& 0.09 & 0.09 &Y  \\
12 &C/D& 18h29m 56.20s & 1$^{\circ}$ 13' 04.1''& 0.08 & 0.08 &Y  \\
13 &C/D& 18h29m 56.53s & 1$^{\circ}$ 13' 11.1''& 0.12 & 0.12 &Y  \\
14 &C/D& 18h29m 57.00s & 1$^{\circ}$ 12' 47.1''& 0.24 & 0.24 &Y  \\
15 &C/D& 18h29m 57.67s & 1$^{\circ}$ 13' 00.2''& 0.10 & 0.10 &Y  \\
16 &C/D& 18h29m 57.71s & 1$^{\circ}$ 14' 06.2''& 0.17 & 0.17 &Y  \\
17 &C/D& 18h29m 57.87s & 1$^{\circ}$ 12' 46.2''& 1.65 & 1.82 &Y  \\
17'&C/D& 18h29m 57.89s & 1$^{\circ}$ 12' 49.7''& 0.18 & 0.18 &Y  \\
18 &C/D& 18h29m 59.64s & 1$^{\circ}$ 14' 12.3''& 1.83 & 1.83 &Y  \\
19 &C/D& 18h30m 00.00s & 1$^{\circ}$ 13' 11.3''& 0.11 & 0.11 &Y  \\
19'&C/D& 18h30m 00.06s & 1$^{\circ}$ 13' 08.3''& 0.08 & 0.08 &Y? \\
20 &C/D& 18h30m 00.65s & 1$^{\circ}$ 13' 40.4''& 0.36 & 0.36 &Y  \\
\hline
\end{tabular}

$^a$Flux density values have been corrected by the the primary beam response 
of the VLA. Sources VLA 1, VLA 2, VLA 3, and VLA 6 are far from the phase 
center, outside the primary beam lobe, with flux density values uncertain. \\
$^b$ Synthesized beam sizes are given in Table 1.\\

\label{radiosources}
\end{table*}


We have detected 22 VLA 3.5 cm continuum sources towards the Serpens
core. Fig ~\ref{svs20_radio} presents 3.5 cm continuum contour maps of the 
SVS 20 and Serpens triple radio source fields. Table \ref{radiosources} 
gives the J2000.0 equatorial coordinates and peak and total flux densities. 
These flux densities have been corrected by the primary beam response of the 
VLA. All the continuum sources are point-like for our synthesized beam sizes, 
with the exception of VLA 7, which is the well known Serpens triple radio 
source (Curiel et al. 1993), and VLA 17, which is associated with the near-IR 
source EC 95 (see also Smith et al. 1999). We also note that SVS 20 (phase 
center of the C/D configuration) and the Serpens thermal radio 
jet (phase center of the D configuration) are separated by 2.4$'$
and therefore both sources are at the border but within the two overlapping 
primary beams of the VLA (5.4' at 3.5 cm). However, we have preferred to manage
the two sets of data (C/D and D configurations) separately, without merging 
them into a single data set, since most of the detected sources are not 
overlapping within these two primary beams (see Fig. 1); in this way, more 
accurate flux density measurements can be achieved. We have nevertheless 
checked the flux density of the compact source VLA 16 (SVS20) as measured 
with the D data set after primary beam correction, finding that it is fully 
consistent (within $\simeq 1 \sigma, \simeq 6 \,\%$) with the flux density 
measured with the C/D configuration.

\begin{table*}
\label{radioassociations}
\caption{Counterparts of VLA radio sources at other wavelengths$^a$}
\begin{tabular}{llllll}
\hline		 		
VLA &Near-IR          &Mid-IR (ISO) &smm/mm & X-Rays & Class\\
\hline			      
1  & 2MASS18293461+0115093?&             &       &        &\\
2  & 2MASS18293461+0115093?&             &       &        &\\
4  & EC 27?           &             &       &        &\\
5  & smm 9-IR         &ISO 241      &smm 9  &        &0,I\\
7  & EC 41            &ISO 258a,258b&smm 1  &        &0,I\\
9  & EC 53            &ISO 265      &smm 5  &1829512+011640&I\\
10 & smm 10-IR        &ISO 270      &smm 10 &        &0,I\\
11 & GCNM 53          &ISO 276      &       &        &I\\
12 & EC 79            &ISO 304      &       &        &II\\
13 &                  &ISO 308?     &smm 4  &        &0,I\\
14 & EC 84            &ISO 309      &       &        &II\\
15 & EC 88            &ISO 312      &       &        &I\\
16 & SVS 20           &ISO 314      &smm 6  &1829577+011407&flat\\
17 & EC 95            &ISO 317$^b$   &       &1829579+011246&II\\
17'& EC 92            &ISO 317$^b$   &       &        &I\\
18 & GCNM 122         &             &       &        &\\
19 &                  &ISO 331      &       &        &I\\
19'&                  &ISO 331?     &       &        &I\\
20 & EC 117           &ISO 338      &       &1830007+011338&II\\
\hline
\end{tabular}

$^a$ Labels in the different columns refer to the following works: EC 
(Eiroa \& Casali, 1992), GCNM (Giovannetti et al. 1998), ISO (Kaas et 
al, 2004), smm (Casali et al. 1993, Davis et al. 1999). Sources in the 
X-Ray column are from Preibisch (2003).
$^b$ Kaas et al. (2004) do not resolve the flat spectrum source ISO 317 and 
associate it with both EC 92 and EC 95. More recently, Pontoppidan et al. 
(2004) clearly resolve the ISO fluxes into two distinct sources; from their 
fluxes we have estimated the evolutionary class (last column).  
\end{table*}


Following the formulation given by Anglada et al. (1998), we have estimated 
that the number of random background 3.5 cm continuum sources expected within 
a field size of 5.4$'$  (primary beam size) with a detection threshold of 
$\simeq$ 0.08 mJy ($\simeq$ 5$\sigma$) is $\simeq$ 1. This means that almost
all the detections within the primary beam of the two sets of data (C/D and D 
configurations) are likely associated with star formation in the Serpens core
and, in fact, this is in good agreement with our findings, as it is shown in
the following. The last column of Table \ref{radiosources} indicates whether 
the radio sources have plausible counterparts in other wavelength ranges (see 
Section A, appendix for a short description of the radio continuum sources and 
their associated counterparts). 16 out of the 22 VLA sources (including in this
number 4 sources outside the 
primary beam, see note in Table \ref{radiosources}) are suggested to be 
associated with near-IR objects, ISO mid-IR sources, mm/sub-mm sources, and 
X-ray sources. All these sources are YSOs embedded in the Serpens core; their 
names together with the VLA names are given in Table \ref{radioassociations}. 
Given the uncertainties in the absolute positions and the radio beam sizes, as 
well as the fact that positions of the near-IR sources can differ 
$\sim 2 - 3$ arcsec in each equatorial coordinate when comparing different 
near-IR works (Eiroa \& Casali 1992, Sogawa et al. 1997, Giovannetti et 
al. 1998, Kaas 1999, Pontoppidan et al. 2004), an association between a radio 
and a near-IR source is done when  their positions coincide within a projected
angular distance of 5 arcsec. This is a conservative estimate given
the uncertainties between the radio and near-IR positional reference frames, 
as well as the near-IR random positional errors (estimated to be $\sim 3 $ 
arcsec, considering the aforementioned differences among
different authors); the relative radio position uncertainties are very low, 
of the order of 0.1 arcsec. Last column of Table \ref{radioassociations} 
gives the usual updated SED classification of YSOs (e.g. Andr\'e \& Montmerle
1994, Greene et al. 1994) of the Serpens radio sources. Some 
objects are marked as belonging to two different evolutionary classes, i.e., 
Class 0/Class I. In these cases, the Class 0 classification is based on their 
submillimeter emission (Casali et al. 1993; Hurt \& Barsony 1996; Wolf-Chase 
et al. 1998; Hogerheidje et al. 1999; Brown et al. 2000), while the Class I 
classification is taken from their ISO mid-IR emission (Kaas et al. 2004).
Davis et al. (1999) suggested that the Serpens protostellar objects are 
borderline objects between Class 0 and Class I protostars. It is worth to note 
that the SED Class 0/Class I ``ambiguity'' has recently been noted in Cep E by
Noriega-Crespo et al. (2004), who suggest that a reacommodation of the Class 
0/Class I SED classification is needed given flux-limit sensitivities of 
current mid-IR observations (e.g. Spitzer). In general, the flat-spectrum and 
Class II classifications in Table ~\ref{radioassociations} are taken 
from ISO fluxes of the objects (Kaas et al. 2004), although we make the 
following considerations. VLA 15 is associated with the near-IR source EC 88 
and with ISO 312. We note that ISO 312 is associated with EC 88 and EC 89 by 
Kaas et al. (2004), because both near-IR sources are not resolved by the ISO 
beam. However, Pontoppidan et al. (2004) have recently separated the mid-IR 
ISO emission from both EC 88 and EC 89 undependently. Using the data of these 
later authors, we classify EC 88 as a Class I protostar, and EC 89 as a flat-spectrum source. A similar case is represented by VLA 17/VlA 17' and the 
flat-spectrum source ISO 317 (Kaas et al. 2004), recently  resolved by 
Pontoppidan et al. (2004) into two ISO sources.  In this case, VLA 17 is 
associated with EC 95, which has a Class II SED (using the fluxes given by 
Pontoppidan et al. 2004), and VLA 17' is associated with EC 92, with a Class I 
SED. Finally, VLA 16 is associated with the binary star SVS 20/EC 90, which is 
ISO 314 (Kaas et al. 2004); in this case, Haisch et al.  (2002) classify both 
components of the binary SVS 20 as flat-spectrum sources on the basis of 
ground-based 10 $\mu$m observations.

Five out of the 16 radio sources associated with near/mid-infrared ones 
(VLA 5, VLA 7, VLA 9, VLA 10, and VLA 11) are found in the NW cloud clump, 
i.e., the FIRS 1/S 68N field located towards the NW of the Serpens reflection 
nebula (SRN), while the rest of the sources (VLA 12, VLA 13, VLA 14, VLA 15, 
VLA 16, VLA 17, VLA 17', VLA 18, VLA 19, VLA 19', and VLA 20) are located in 
the Southeastern (SE) cloud clump (South from SRN), extending from SVS 20 
towards the SVS 4 complex.  

With respect to the other detected radio sources, VLA 3, VLA 6, and VLA 8 have 
no identifications at other wavelength regimes and they most likely are 
background extragalactic objects. On the other hand, VLA~1 and VLA~2 have 
dubious associations. These two sources are located close to the radio source 
S 68/5 detected at 1.4 GHz by Snell \& Bally (1986). It could be that S 68/5 
was not resolved at 1.4 GHz into our two sources since it approximately lies 
in the middle of VLA 1 and 2. By adding the integrated flux densities of VLA 1 
and VLA 2 and comparing them with the 1.4 GHz flux density of S 68/5, we obtain
a negative spectral index suggesting that it could be a 
background extragalactic object. A faint 2MASS source, 18293641+0115093, 
only detected at $J$ (17.00 mag) and $H$ (15.74 mag) is located at a  
distance of $\simeq$ 3$''$ of either VLA 1 and 2. Finally, the weak radio 
source VLA 4 is at a projected distance of $\sim$ 6'' from the near-IR source 
EC 27 ($K$ = 13.7 mag), but we think they are unrelated. Both radio and 
near-IR sources are  $\sim$ 25'' towards the Northwest of S 68N. 

\section{Discussion}
\label{Discussion}

\subsection{Nature of the radio sources}

Centimeter continuum emission has been detected towards a large number of 
low luminosity YSOs, usually powering outflows and most of them in the Class 
0/Class I protostellar phase, but also few in the Class II phase (embedded 
T Tauri stars). That emission is very weak with flux density values 
in the range of tenths to a few mJy. The emission is usually compact, although 
when observed at subarcsecond resolution, the sources tend to present a 
jet-like structure (e.g., Reipurth et al. 2004). Measurements of spectral 
indices at cm wavelengths reveal that thermal free-free 
emission dominates at these frequencies, although there exists a non-thermal 
component in few cases, as it is the case of the Serpens radio jet - our source
VLA 7 (Curiel et al. 1993). Theoretical approaches explaining the origin of the
radio emission usually invoke winds (disk-winds or X-winds) from an embedded 
protostellar object since stellar radiation produces too little ionization to 
account for the continuum observations (Torrelles et al. 1985; Curiel et al. 
1989; Shang et al. 2004, and references therein). On the other hand, radio 
counterparts of X-ray YSOs have also been explained in terms of non-thermal 
gyrosynchroton emission arising from coronally active stars; in fact, this is 
the suggestion made by Smith et al. (1999) for VLA 17 (EC 95), which is 
associated with a very strong X-ray source (Preibisch 1998, 2003). 
We note that the  Serpens VLA sources follow the general properties just 
described, although we have no information on their radio continuum SED and 
sizes at the subarcsecond scale. 

The Serpens radio sources are associated with YSOs in different evolutionary 
stages, i.e., there are Class 0, Class I, flat-spectrum and Class II objects,
but the trend is to be associated with the youngest Class 0/Class I 
protostars (last column of Table ~\ref{radioassociations}). Some of the 
Class 0 sources might present observational evidences of infalling gas, as 
described in the appendix. This variety of evolutionary phases might be the 
cause why there is no correlation of the radio flux density neither with the 
observed K magnitudes, nor with mid-IR fluxes, nor with SED IR spectral 
indexes. Radio molecular observations reveal a widespread, very complex 
outflow activity in the Serpens cloud core, which makes it complicated to 
identify individual molecular outflows (McMullin et al. 1994; White et al. 
1995, Davis et al. 1999). Nevertheless, some of the radio sources can be 
associated with relatively
well defined individual outflows: VLA 7 is the Serpens radio jet; VLA 5, VLA 
10, VLA 11, and VLA 13 are associated with molecular and/or H$_2$ outflows. 
All of them are Class 0/Class I objects. Further, VLA 9 is a Class I 
object associated with the cometary nebulosity EC 53; VLA 18 and VLA 20 also 
shows some indications of nebulosity with a cometary morphology (very faint in 
the case of VLA 18). This is suggestive of outflows in these sources. Thus, 
it is likely that most of the Serpens radio sources are YSO thermal radio jets 
as those found associated with many other protostellar objects 
(Reipurth et al. 2004), at least in the cases of VLA 5, VLA 7, VLA  10, 
VLA 11, and VLA 13. 

On the other hand, VLA 9, VLA 16, VLA 17, and VLA 20 are associated with X-ray 
sources. The radio emission of the Class II source VLA 17 (EC 95) has been 
explained by Smith et al. (1999) as due to gyrosynchrotron emission in an 
extreme stellar corona. VLA 9 (EC 53) 
is a Class I object, but with very deep CO absorption bands in its K-spectrum 
and EXOR-type properties; thus a more evolved Class II nature can not be ruled 
out. VLA 16 is associated with one of the components of the flat-spectrum 
binary source SVS 20, most likely the Northern component, while VLA 20 is a 
Class II object. Interestingly, the X-ray/radio luminosity ratio of all Serpens
radio sources with X-ray emission approximately follows the ratio of coronal 
active stars (G\"udel \& Benz 1993, Smith et al. 1999), which suggests that 
these Serpens radio sources could be examples of YSOs with that kind of 
activity.

\subsection{Clustering}

The Serpens VLA sources can be grouped into two groups. As mentioned above 
(\S 3), the first group (VLA 5, VLA 7, VLA 9, VLA 10, and VLA 11) are located 
towards the NW clump of the Serpens cloud core. The rest of the sources form a 
second group located in the SE clump of the Serpens core (Table 
\ref{radioassociations} and Fig. ~\ref{svs20_radio}). Both core clumps are 
well known from a variety of molecules, like NH$_3$, C$^{18}$0, NH$^+$ or CS 
(Ungerechts \& G\"usten 1984; White et al. 1995; McMullin et al. 2000; Olmo 
\& Testi 2002). The radio sources in the NW condensation are all associated 
with very young Class 0/Class I protostars, while at least half of the radio 
sources in the SE clump are associated with more evolved flat-spectrum and 
Class II sources. This result is coherent with the picture of star forming 
activitiy emerging from results in other wavelengths. Testi et al. (2000) 
already noticed subclustering of YSOs within both clumps of the Serpens 
mm/sub-mm sources. Kaas et al (2004) find a similar distribution  for the ISO 
Class I sources, while their Class II sources are more widespread distributed, 
although with a trend to be more concentrated in the SE condensation, which is 
also the case of the near-IR cluster (Eiroa \& Casali 1992; Giovannetti et al. 
1998; Kaas 1999). It is most likely that we observe radio emission in the NW 
clump only from Class 0/Class I sources due to the scarcity of more evolved 
stars in that part of the cloud. Thus, the data suggest that although stars are
presently forming in both NW and SE cloud condensations, it began earlier or it
was more efficient in the SE clump. An interesting case is that of the stellar 
group SVS 4. From the $\sim$ 10 stars forming the SVS 4 subcluster and sharing
a common near-IR nebulosity, 5 of them are most likely associated with radio 
sources. Two of these sources, VLA 15 (EC 88) and VLA 17' (EC 92) are 
classified as Class I objects; the rest are more evolved Class II stars.

\section{Conclusions}
\label{Conclusions}

The Serpens core is known to host a rich YSO cluster with more than 150 
members, including a large fraction of embedded very young Class 0/Class 
I protostars. Our VLA 3.5 cm observations reveal a number of radio continuum 
sources associated with Serpens YSOs and provide a first insight into the 
radio continuum properties of the young cluster. Most of these radio sources, 
10 out of 16, are associated with Class 0/Class I objects, one is a 
flat-spectrum source, and  only 4 out of 16 are associated with more evolved
 Class II (embedded Classical T Tauri stars) sources. The source VLA 18 (GCNM 
122) has not been classified in terms of evolutionary classes. At first sight
the ratio between Class 0/Class I and flat-spectrum/Class II sources detected 
in radio continuum seems to be in contrast with the YSO population found at 
infrared wavelengths, where Kaas et al. (2004) identified 20 Class I, 13 
flat-spectrum and 43 Class II sources. This result plausibly reflects the fact
that radio continuum sources tend to be preferentially associated with the 
earliest phases of star formation, rather than with the more evolved T Tauri 
phase. The data do not allow us to trace any firm conclusion on the nature of 
the radio emission. We tentatively suggest that the emission from Class 
0/Class I sources is due to thermal radio jets, as suggested by Anglada et al.
(1998) and Reipurth et al. (2004) for similar sources, while the radio 
emission from the Serpens Class II sources could have a stellar coronal 
origin, as suggested by Smith et al. (1999) for EC 95. A determination of 
radio spectral indices is needed in order to ellucidate between both or any 
other different alternatives, which requires simultaneous observations at 
different radio frequencies.

\acknowledgments
CE and JMT acknowledge partial finantial support from Spanish grants 
PNAYA 2001-1124 and PNAYA 2002-00376, respectively. SC is grateful to CONACyT, 
M\'exico and DGAPA, UNAM for their finantial support. The comments of an 
anonymous referee helped us to improve the content of the paper.



\appendix

\section{Appendix: comments on individual sources}
\label{Individual}

\subsection{VLA5}

VLA 5 is the Class 0 source S 68N/smm 9 and ISO 241 (Casali et al. 1993, 
McMullin et al. 1994, White et al. 1995, Davis et al. 1999, Kaas et al. 
2004). H$_2$O maser emission and compact molecular outflows in several lines,
e.g., CO, CS are observed (Wolf-Chase et al. 1998,  Williams \& Myers
2000). Infalling gas in H$_2$CO and CS lines has been suggested (Hurt et 
al. 1996, Mardones et al. 1997, Williams \& Myers 2000). Hodapp (1999) 
noticed the presence of a very red, near-IR double star, smm 9-IR, associated 
with some nebulosity near the nominal position of the smm source (see the 
excellent Figs. 1 and 2 in Hodapp's paper); he interestingly pointed out the 
intriguing possibility of an association of the Class 0 source with one of the
near-IR objects. Table ~\ref{VLA5} lists the positions of the objects 
associated with VLA 5. The position of smm 9-IR has been estimated by us 
using the $K$ image of Kaas (1999). 

\begin{table}[h]
\caption{Equatorial coordinates of VLA 5 and associated objects}
\begin{tabular}{lcl}
\hline
Object& $\alpha ~(2000.0)~ \delta$        &Ref\\
\hline
VLA 5          & 18h29m48.10s  1$^{\circ}$16$'$44.7$''$ &1\\
smm 9-IR       & 18h29m48.03s  1$^{\circ}$16$'$44.4$''$ &1\\
ISO 241        & 18h29m48.1s~  1$^{\circ}$16$'$43$''$~~ &2\\
smm 9          & 18h29m48.10s  1$^{\circ}$16$'$41.5$''$ &3\\
smm 9 (H$_2$0) & 18h29m48.02s  1$^{\circ}$16$'$43.5$''$ &4\\
\hline
\end{tabular}

(1) This work, (2) Kaas et al. (2004),
(3) Davis et al. (1999), (4) Wolf-Chase et al. (1998)
\label{VLA5}
\end{table}

\subsection{VLA7}

VLA 7 is the well studied  triple radio-jet associated with the most prominent 
Class 0 object in Serpens, FIRS 1/smm 1 (Curiel et al. 1993, Harvey et al. 
1984, Casali et al. 1993, Hurt \& Barsony 1996). H$_2$O maser emission, 
molecular outflows and a molecular envelope are also found (e.g. Dinger \& 
Dickinson 1980, White et al. 1995, Curiel et al. 1996, Hogerheijde et al. 1999,
Williams \& Myers 2000). An H$_2$ near-IR jet with a point-like source at its 
apex, EC 41, is observed close to the position of the radio/smm source. Both 
objects likely are  different YSOs (Eiroa \& Casali 1989, Hodapp 1999). Kaas 
et al.(2004) have detected two mid-IR sources in this region: ISO 258a, a 
Class I source likely associated with EC 41, and ISO 258b, which could be 
associated with VLA 7/smm 1, although Kaas et al. (2004) find it unlikely. 
Table \ref{VLA7} lists the positions of the sources.  

\begin{table}[h]
\caption{Equatorial coordinates of VLA 7 and associated objects}
\begin{tabular}{lcl}
\hline
Object& $\alpha ~(2000.0)~ \delta$        &Ref\\
\hline
VLA 7          & 18h29m49.79s 1$^{\circ}$15$'$20.8$''$ &1\\
smm 1          & 18h29m49.8s  ~1$^{\circ}$15$'$18.6$''$ &2\\
EC 41          & 18h29m49.68s 1$^{\circ}$15$'$28.6$''$&3\\
ISO 258a       & 18h29m49.6s ~1$^{\circ}$15$'$28$''$~~ &4\\
ISO 258b       & 18h29m50.3s ~1$^{\circ}$15$'$21$''$~~  &4 \\
\hline
\end{tabular}

(1) This work, (2) Davis et al. (1999), (3) Eiroa \& Casali (1992),
(4) Kaas et al. (2004)
\label{VLA7}
\end{table}

\subsection{VLA 9}

VLA 9 is associated with the variable near-IR reflection nebulosity EC 53 and 
the Class I source smm 5/ISO 265 (Eiroa \& Casali 1992, Casali et al. 1993, 
Hodapp et al. 1996, Horrobin et al. 1997, Sogawa et al. 1997, Kaas 1999, Davis 
et al. 1999, Kaas et al. 2004). There is no near-IR point-like source 
embedded in the nebulosity; Hodapp (1999) suggested an EXor-type object 
nature for the embedded stellar object in EC 53. The $K$-spectrum presents 
strong CO absorption bands (Casali \& Eiroa, 1996). There is no prominent 
molecular line emission although marginal blueshifted CO emission close to 
smm 5 might be present (White et al. 1995, McMullin et al. 2000, Williams \& 
Myers 2000, Narayanan et al. 2002). An X-ray source is associated with EC 53 
(Preibisch 2003, 2004). VLA 9 likely denotes the position of the EXor-type 
star in EC 53 (Table \ref{VLA9}).  

\begin{table}[h]
\caption{Equatorial coordinates of VLA 9 and associated objects}
\begin{tabular}{lcl}
\hline
Object& $\alpha ~(2000.0)~ \delta$        &Ref\\
\hline
VLA 9    & 18h29m51.30s 1$^{\circ}$16$'$40.9$''$ &1 \\
EC 53    & 18h29m51.17s 1$^{\circ}$16$'$40.7$''$ &2 \\
ISO 265  & 18h29m51.2s~ 1$^{\circ}$16$'$42$''$~~ &3\\
smm 5    & 18h29m51.10s 1$^{\circ}$16$'$35.7$''$ &4\\
182951.2+011640&18h29m51.2s~~ 1$^{\circ}$16$'$40$''$~~~&5 \\  
\hline
\end{tabular}

(1) This work, (2) Eiroa \& Casali (1992), (3) Kaas et al. 
(2004), (4) Davis et al. (1999), (5) Preibisch (2003)

\label{VLA9}
\end{table}

\subsection{VLA 10}

\begin{table}[h]
\caption{Equatorial coordinates of VLA 10 and associated objects}
\begin{tabular}{lcl}
\hline
Object& $\alpha ~(2000.0)~ \delta$        &Ref\\
\hline
VLA 10    & 18h29m52.12s 1$^{\circ}$15$'$46.0$''$ &1\\
smm 10-IR & 18h29m52.18s 1$^{\circ}$15$'$47.8$''$ &1\\
ISO 270   & 18h29m52.2s~ 1$^{\circ}$15$'$49$''$~~ &2\\
smm 10    & 18h29m52.10s 1$^{\circ}$15$'$47.7$''$ &3\\

\hline
\end{tabular}

(1) This work, (2) Kaas et al. (2004), (3) Davis et al. (1999)
\label{VLA10}
\end{table}

VLA 10 coincides with the Class 0 source smm 10/ISO 270 (Davis et al. 
1999, Testi et al. 1998, Kaas et al. 2004) and with a very red, highly 
variable near-IR object (Hodapp 1999). The position of smm 10-IR has 
been estimated by us using the $K$ image of Kaas (1999). It might be a 
second example of a near-IR object associated with a Class 0 object 
(Table ~\ref{VLA10}). A CS weak outflow is detected by Williams \& Myers 
(2000).

\subsection{VLA 11}

VLA 11 is associated with the near-IR source GCNM 53 and the Class I source 
ISO 276, just located towards the Northwest of SRN, Table \ref{VLA11} 
(Giovannetti et al. 1997, Kaas et al. 2004). The source is at the basis of
a H$_2$ jet oriented in the North-West direction with a position angle of 
around 135 degrees (Herbst et al. 1997). 

\begin{table}[h]
\caption{Equatorial coordinates of VLA 11  and associated objects 
objects}
\begin{tabular}{lcl}
\hline
Object& $\alpha ~(2000.0)~ \delta$        &Ref\\
\hline
VLA 11    & 18h29m52.94s 1$^{\circ}$14$'$54.0$''$ &1\\
GCNM 53   & 18h29m52.80s 1$^{\circ}$14$'$54.8$''$ &2\\
ISO 276   & 18h29m52.9s~ 1$^{\circ}$14$'$56$''$~~ &3\\

\hline
\end{tabular}

(1) This work, (2) Giovannetti et al. (1998), (3) Kaas et al. (2004)
\label{VLA11}
\end{table}

\subsection{VLA 12}

\begin{table}[h]
\caption{Equatorial coordinates of VLA 12  and associated objects}
\begin{tabular}{lcl}
\hline
Object& $\alpha ~(2000.0)~ \delta$        &Ref\\
\hline
VLA 12    & 18h29m56.20s 1$^{\circ}$13$'$04.1$''$ &1\\
EC 79     & 18h29m56.54s 1$^{\circ}$13$'$01.1$''$ &2 \\
ISO 304   & 18h29m56.6 ~ 1$^{\circ}$13$'$01$''$~~ &3\\
\hline
\end{tabular}

(1) This work, (2) Eiroa\& Casali (1992), Kaas et al. (2004)
\label{VLA12}
\end{table}

VLA 12 is at a angular distance of $\simeq$ 5$''$ of the Class II source 
EC 79/ISO 304 (Table ~\ref{VLA12}). EC 79 is star 2 of the SVS 4 complex. 
The spectral type of  EC 79 is late K, M as suggested by NaI, CaI and CO 
absorptions present in its $K$ spectrum (Eiroa \& Djupvik 2005, in 
preparation). We tentatively associate the VLA source with the Class II/T 
Tauri-like YSO.

\subsection{VLA 13}

VLA 13 coincides with the Class 0 object smm 4 and maybe with ISO 308 
(Table \ref{VLA13}); this protostellar object shows evidences of infalling gas 
and molecular outflows (Casali et al. 1993, White et al. 1995, Hurt et al. 
1996, Mardones et al. 1997, Gregersen et al. 1997, Davis et al. 1999, 
Hogerheidijde et al. 1999, Larson et al. 2000, Narayanan et al. 2002, Kaas et 
al. 2004). Two compact $H_2$ condensations (Eiroa et al. 1997, Herbst et al. 
1997) are located close to smm 4 and likely related to it. 

\begin{table}[h]
\caption{Equatorial coordinates of VLA 13 and associated objects}
\begin{tabular}{lcl}
\hline
Object& $\alpha ~(2000.0)~ \delta$        &Ref\\
\hline
VLA 13    & 18h29m56.53s 1$^{\circ}$13$'$11.1$''$ &1 \\
smm 4     & 18h29m56.50s 1$^{\circ}$13$'$10.1$''$ &2 \\
ISO 308   & 18h29m56.80s 1$^{\circ}$13$'$18$''$ ~~  & 3\\ 
\hline
\end{tabular}
\label{VLA13}

(1) This work, (2) Davis et al. (1999), Kaas et al. (2004)
\end{table}

\subsection{VLA 14}

VLA 14 is associated with the Class II source EC 84/ ISO 309, a member of the 
SVS 4 stellar subgroup (Table \ref{VLA14}). NaI and CO absorptions are detected
in the $K$ spectrum of this source (Eiroa \& Djupvik 2005, in preparation).

\begin{table}[h]
\caption{Equatorial coordinates of VLA 14 and associated objects}
\begin{tabular}{lcl}
\hline
Object& $\alpha ~(2000.0)~ \delta$        &Ref\\
\hline
VLA 14    & 18h29m57.00s 1$^{\circ}$12$'$47.1$''$ &1\\
EC 84     & 18h29m57.00s 1$^{\circ}$12$'$47.6$''$ &2 \\
ISO 309   & 18h29m56.80s 1$^{\circ}$12$'$49$''$~~ &3\\ 
\hline
\end{tabular}

(1) This work, (2) Pontoppidan et al. (2004), (3) Kaas et al. (2004) 
\label{VLA14}
\end{table}

\subsection{VLA 15}

VLA 15 is most likely associated with the variable  star EC 88 
(Eiroa \& Casali 1992, Kaas 1999). EC 89 is found very close to it and has 
faint Br$_\gamma$ emission (Eiroa \&  Djupvik 2005, in preparation). Both 
belong to the SVS 4 stellar group. Kaas et al. (2004) associate the Class I 
object ISO 312 with both EC 88/EC 89 stars, but using the results of 
Pontoppidan et al. (2004) EC 88 (VLA 15) is a Class I object, 
while EC 89 is a flat-spectrum source (Table \ref{VLA15}).

\begin{table}[h]
\caption{Equatorial coordinates of VLA 15 and associated objects}
\begin{tabular}{lcl}
\hline
Object& $\alpha ~(2000.0)~ \delta$        &Ref\\
\hline
VLA 15    & 18h29m57.67s 1$^{\circ}$13'00.1'' &1\\
EC 88     & 18h29m57.63s 1$^{\circ}$13'00.2'' &2 \\
EC 89     & 18h29m57.69s 1$^{\circ}$13'04.4'' &2 \\
ISO 312   & 18h29m57.50s 1$^{\circ}$12'59'' ~~&3 \\
\hline
\end{tabular}

(1) This work, (2) Pontoppidan et al. (2004), Kaas et al. (2004)
\label{VLA15}
\end{table}

 \subsection{VLA 16}

\begin{table}[h]
\caption{Equatorial coordinates of VLA 16 and associated objects}
\begin{tabular}{lcl}
\hline
Object& $\alpha ~(2000.0)~ \delta$        &Ref\\
\hline
VLA 16    & 18h29m57.71s 1$^{\circ}$14'06.1'' &1\\
SVS 20-S  & 18h29m57.62s 1$^{\circ}$14'03.1'' &2 \\
SVS 20-N  & 18h29m57.69s 1$^{\circ}$14'04.8'' &2 \\
smm 6     & 18h29m57.60s 1$^{\circ}$14'02.1'' &3 \\
ISO 314   & 18h29m57.50s 1$^{\circ}$14'07''~~ &4\\
182957.7+011404& 18h29m57.70s 1$^{\circ}$14'04''~~&5\\
\hline
\end{tabular}
\label{VLA16}

(1) This work, (2) Eiroa et al (1987), (3) Davis et al (1999), (4)
Kaas et al. (2004), (5) Preibisch (2003) 
\end{table}

VLA 16 is spatially associated with smm 6, ISO 314, an X-ray source and the 
near-IR binary SVS 20/EC 90, the strongest near-IR source embedded in the 
Serpens cloud core (Eiroa et al. 1987, Eiroa \& Casali 1992, Casali et al.
1993, Davis et al. 1999, Preibisch 2003, Kaas et al. 2004); both components of 
the binary present flat-spectrum SEDs (Haisch et al. 2002). The source is 
detected at 8.4 GHz by Smith et al. (1999). The near-IR binary is 
surrounded by a ring of nebulosity with two spiral arms seen in opposite 
directions; $H_2$ emission is detected at the end of the SW arm. The ring could
be a circumbinary disk (Eiroa et al. 1997a); alternatively, Huard et al. 
(1997) suggest that the ring could delineate evacuated, bipolar cavities. The 
brightest component of the binary, SVS 20S, presents a featureless $K$ 
spectrum, while SVS 20N has strong Br$_\gamma$ and CO bands in emission (Herbst
\& Rayner 1994, Eiroa \&  Djupvik 2005, in preparation), indicating the presence of
an accretion disk. It might be that VLA 16 is associated with SVS 20N (Table 
\ref{VLA16}). 

\subsection{VLA 17 and VLA17'}

VLA 17 is associated with EC 95, a young  star with strong X-ray emission, 
while VLA 17' is  most likely related to the nearby star EC 92  (Table 
~\ref{VLA17}). EC 95 and EC 92 belong to the SVS 4 group and present NaI, CaI 
and CO absorptions  in their $K$ spectra (Preibisch  1999, Eiroa \& Djupvik 
2005, in preparation).  The flat-spectrum source ISO 317 corresponds to both 
stars (Kaas et al. 2004), but it is resolved by Pontoppidan et al (2004) into 
two distinct mid-IR sources: a Class II source (EC 95) and a Class I source 
(EC 92). Smith et al. (1999) suggest optically thin synchrotron emission as 
the origin of the VLA 17 (EC 95) radio continuum emission.
 
\begin{table}[h]
\caption{Equatorial coordinates of VLA 17, VLA 17' and associated 
objects}
\begin{tabular}{lcl}
\hline
Object& $\alpha ~(2000.0)~ \delta$        &Ref\\
\hline
VLA 17    & 18h29m57.87s 1$^{\circ}$12'46.2'' &1\\
VLA 17'   & 18h29m57.89s 1$^{\circ}$12'49.7'' &1\\
EC 95     & 18h29m57.92s 1$^{\circ}$12'46.0'' &2\\
EC 92     & 18h29m57.88s 1$^{\circ}$12'51.1'' &2\\
ISO 317   & 18h29m57.80s 1$^{\circ}$12'52''~~   &3\\
1829579+011246&18h29m57.90s 1$^{\circ}$12'46''~~ &4\\ 
\hline
\end{tabular}

(1) This work, (2) Pontoppidan et al. (2004), (3) Kaas et al. (2004)
(4) Preibisch (2003)

\label{VLA17}
\end{table}

\subsection{VLA 18}

VLA 18 is spatially associated with the weak near-IR source GCNM 122 
(Table ~\ref{VLA18}), which shows a very faint cometary-like nebulosity. The 
radio source was also detected at 8.4 GHz by Smith et al. (1999). GCNM 122 is 
variable: Giovannetti et al. (1998) gives a $K$ magnitude of 14.9, while we 
estimate 16.3 mag. 

\begin{table}[h]
\caption{Equatorial coordinates of VLA 18 and associated objects}
\begin{tabular}{lcl}
\hline
Object& $\alpha ~(2000.0)~ \delta$        &Ref\\
\hline
VLA 18   & 18h29m59.64s  1$^{\circ}$14'12.3'' &1\\
GCNM 122 & 18h29m59.51s  1$^{\circ}$14'12.3''&2\\
\hline
\end{tabular}

(1) This work, (2) Giovannetti et al. (1998)
\label{VLA18}
\end{table}

\subsection{VLA 19 and VLA 19'}

The pair of weak radio sources VLA 19 and VLA 19' are likely associated 
with the Class I source ISO 331 (Table \ref{VLA19}), but have no near-IR 
counterparts. VLA 19 was detected also at 8.4 GHz by Smith et al. (1999). 

\begin{table}[h]
\caption{Equatorial coordinates of VLA 19, VLA 19' and associated 
objects}
\begin{tabular}{lcl}
\hline
Object& $\alpha ~(2000.0)~ \delta$        &Ref\\
\hline
VLA 19   & 18h30m00.00s  1$^{\circ}$13'11.3'' &1\\
VLA 19'  & 18h30m00.06s  1$^{\circ}$13'08.3'' &1\\
ISO 331  & 18h29m59.7s  ~1$^{\circ}$13'13'' ~~&2\\
\hline
\end{tabular}

(1) This work, (2) Kaas et al. (2004)
\label{VLA19}
\end{table}

\subsection{VLA 20}

VLA 20 is associated with the Class II object EC 117/ISO 338 and a X-ray 
source (Table \ref{VLA20}). The source was detected at 8.4 GHz by Smith et al. 
(1999). There are NaI, CaI, and CO absorptions in the $K$ spectrum of the 
source (Eiroa \& Kaas 2005, in preparation). A nearby 3 mm source ($\sim$ 4 
arcsec southwards) has been detected by Testi \& Sargent (1998).  

\begin{table}[h]
\caption{Equatorial coordinates of VLA 20 and associated objects}
\begin{tabular}{lcl}
\hline
Object& $\alpha ~(2000.0)~ \delta$        &Ref\\
\hline
VLA 20   & 18h30m0.65s  1$^{\circ}$13'40.4'' &1\\
EC 117   & 18h30m0.52s  1$^{\circ}$13'40.4'' &2\\
ISO 338  & 18h30m0.7s  ~1$^{\circ}$13'38'' ~~&3\\
183000.7+011338&18h30m0.7s ~1$^{\circ}$13'38''~~&4\\
\hline
\end{tabular}

(1) This work, (2) Eiroa \& Casali (1992), (3) Kaas et al. (2004),
(4)Preibisch (2003) 
\label{VLA20}
\end{table}


\begin{thebibliography}{}

\bibitem[]{1147} Andre, P., Montmerle, T. 1994, ApJ 420, 837
\bibitem[]{1099}
Anglada, G., Villuendas, E., Estalella, R. et al. 1998, AJ 116, 2953
\bibitem[]{1101}
Brown, D.W., Chandler, C.J., Carlstrom, J.E., et al. 2000, MNRAS 319, 154
\bibitem[]{1107}
Casali, M.M., Eiroa, C. 1996, A\&A 306, 427
\bibitem[]{1109}
Casali, M.M., Eiroa, C., Duncan, W. D. 1993, A\&A 275, 195
\bibitem[]{1111} 
Curiel, S., Rodr\'{\i}guez, L.F., Cant\'o, J., Bohigas, J.,
Roth, M., Torrelles, J.M. 1989, Astrophysical Letters and Communications, 27,
299
\bibitem[]{1115}
Curiel, S., Rodr\'{\i}guez, L.F., G\'omez, J.F., et al. 1996, ApJ 456, 677
\bibitem[]{1117} 
Curiel, S., Rodr\'{\i}guez, L.F., Moran, J.M., Cant\'o, J. 1993, ApJ 415, 191
\bibitem[]{1119}
Davis, C.J., Matthews, H.E., Ray, T.P., Dent, W.R.F., Richer, J.S. 1999, 
MNRAS 309, 141
\bibitem[]{1122}
Dinger, A.S. C., Dickinson, D.F. 1980 AJ 85, 1247
\bibitem[]{1124}
Eiroa, C., Casali, M.M. 1989, A\&A 223. L17 
\bibitem[]{1126}
Eiroa, C., Casali, M. m. 1992, A\&A 262, 468
\bibitem[]{1128}
Eiroa, C., Lenzen, R., Leinert, Ch., Hodapp, K. -W 1987, A\&A 179, 171
\bibitem[]{1130}
Eiroa, C., Palacios, J., Casali, M. M. 1997a, In: Planets beyond the Solar 
System and the Next Generation of space Missions, ed. D. R. Soderblom, 
ASP Conf. Series 119, p. 107
\bibitem[]{1134}
Eiroa, C., Palacios, J., Eisl\"offel, J., Casali, M.M., Curiel, S. 1997b,
In: Low Mass Star Formation - from Infall to Outflow, eds., Malbet, F. \& 
Castets, A., p.103
\bibitem[]{1138} 
Giovannetti, P., Caux, E., Nadeau, D., Monin, J.-L. 1998, A\&A 330, 990
\bibitem[]{1189} Greene, T.P., Wilking, B.A., Andre, P., Young, E.T., Lada, 
C.J. 1994, ApJ 434, 614
\bibitem[]{1140}
Gregersen, E.M., Evans II, N.J., Zhou, S., Choi, M. 1997, ApJ 484, 256
\bibitem[]{1142}
G\"udel, M., Benz, A. O. 1993, ApJ 405, L63
\bibitem[]{1144}
Haisch, K.E., Barsony, M., Greene, T.P., Ressler, M.E. 2002, AJ 124, 4281
\bibitem[]{1146} 
Harvey, P.M., Wilking, B.A., Joy, M. 1984, ApJ 278, 156
\bibitem[]{1148}
Herbst, T.M., Beckwith, S.V.W., Robberto, M. 1997, ApJ 486, L59
\bibitem[]{1150}
Herbst, T.M., Rayner, J.T. 1994, In: Infrared Astronomy with Arrays, ed.
I. McLean, Kluwer, p. 515
\bibitem[]{1153}
Hodapp, K.-W. 1999, AJ 118, 1338
\bibitem[]{1155}
Hodapp, K.-W., Hora, J.L., Raymer, J.T., Pickles, A.J., Ladd, E.F. 1996,
ApJ 468, 861
\bibitem[]{1158}
Hogerheijde, M.R., van Dishoek, E.W., Salverda, J. M., Blake, G. A. 1999,
ApJ 513, 350
\bibitem[]{1161}
Horrobin, M.J., Casali, M.M., Eiroa, C. 1997 A\&A 320, L41
\bibitem[]{1163}
Hurt, R.L., Barsony, M. 1996, ApJ 460, L45 
\bibitem[]{1165}
Hurt, R.L., Barsony, M., Wooten, A.H. 1996, ApJ 456, 686
\bibitem[]{1167}
Kaas, A.A. 1999, AJ 118, 558
\bibitem[]{1169}
Kaas, A.A., Olofsson, G., Bontemps S., et al. 2004, A\&A 421, 623
\bibitem[]{1173}
Larson, B., Liseau, R., Men'sshchikov, A.B., et al. 2000 A\&A 363, 253 
\bibitem[]{1175}
Lodieu, N., Caux, E. Monin, J.-L., Klotz, A. 2002, A\&A 383, L15
\bibitem[]{1177}
Mardones, D., Myers, P.C., Tafalla, M., et al. 1997, ApJ 489, 719
\bibitem[]{1179}
McMullin, J.P., Mundy, L.G., Blake, G.A., et al. 2000, ApJ 536, 845
\bibitem[]{1181}
McMullin, J.P., Mundy, L.G., Wilking, B.A., Hezel, T., Blake, G.A.
1994, ApJ 424, 222
\bibitem[]{1184}
Narayanan, G., Moriarty-Schieven, G., Walker, C.K., Butner, H. M. 2002
ApJ 565, 319
\bibitem[]{1187}
Noriega-Crespo, A., Moro-Mart\'\i n, A., Carey, S. et al. 2004, ApJS 154, 402
\bibitem[]{1189}
Olmi, L., Testi, L. 2002, A\&A 392, 1052
\bibitem[]{1242} Pontoppidan, K.M., van Dishoek, E.F., Dartois, E. 2004, 
A\&A 426, 925
\bibitem[]{1191}
Preibisch, T. 1998, A\&A 338, L25 
\bibitem[]{1193}
Preibisch, T. 1999, A\&A 345, 583
\bibitem[]{1195}
Preibisch, T. 2003, A\&A 410, 951
\bibitem[]{1250} Reipurth, B., Rodr\'\i guez, L.F., Anglada, G., Bally, J. 2004, 
AJ 127, 1736
\bibitem[]{1197}
Rodr\'\i guez, L.F., Curiel, S., Moran, J.M. et al. 1989, ApJ 346, L85
\bibitem[]{1199}
Rodr\'\i guez, L.F., Moran, J.M., Gottlieb, E.W., Ho, P.T.P. 1980, ApJ
235, 845
\bibitem[]{1202}
Shang, H., Lizano, S., Glassgold, A., Shu, F. 2004, ApJ 612, L69
\bibitem[]{1204}
Smith, K., G\"udel, M., Benz, A.O. 1999, A\&A 349, 475
\bibitem[]{1206}
Snell, R.L., Bally, J. 1986, ApJ 303, 683
\bibitem[]{1208}
Sogawa, H., Tamura, M., Gatley, I., Merrill, K.M. 1997, AJ 113, 1057
\bibitem[]{1210} 
Strom, S.E., Vrba, F., Strom, K.M. 1976, AJ 81, 314
\bibitem[]{1212}
Testi, L., Sargent, A.I. 1998, ApJ 508, L91
\bibitem[]{1214} 
Testi, L., Sargent, A.I., Olmi, l., Onello, J.S. 2000, ApJ 540, L53
\bibitem[]{1216} 
Torrelles, J.M., Ho, P.T.P., Rodr\'{\i}guez, L.F., Cant\'o, J.
1985, ApJ, 288, 595
\bibitem[]{1219}
Ungerechts, H., G\"usten, R. 1984, A\& A 131, 177
\bibitem[]{1221}
White, G.J., Casali, M.M., Eiroa, C. 1995, A\&A 298, 594
\bibitem[]{1225}
Williams, J.P., Myers, P.C. 2000, ApJ 537, 891
\bibitem[]{1227}
Wolf-Chase, G.A., Barsony, M., Wootten H.A., et al. 1998, ApJ501, L193
\bibitem[]{1229}
Ziener, R., Eisl\"offel, J. 1999, A\&A 347, 565

\end{thebibliography}
\end{document}